\documentclass[final,5p,times,twocolumn]{elsarticle}

\usepackage{hyperref}
\hypersetup{
 setpagesize=false,
 colorlinks=true,
 linkcolor=blue,
 citecolor=blue,
 filecolor=blue,
 urlcolor=blue,
}

\usepackage{amssymb}
\usepackage{lipsum}
\usepackage{amsthm}
\usepackage{amsmath}
\usepackage{lineno}

\usepackage{comment}
\usepackage{multirow}

\journal{Physics Letters B}

\begin{document}

\begin{frontmatter}

\title{Measurement of $^{3,4}$He($K^-, \pi^0$)$^{3,4}_\Lambda$H reaction cross section and evaluation of hypertriton binding energy}


\author[12]{T. Akaishi}
\ead{akaishi@rcnp.osaka-u.ac.jp}
\author[2]{H. Asano}
\author[3]{X. Chen}
\author[4]{A. Clozza}
\author[4]{C. Curceanu}
\author[4,14]{R. Del Grande}
\author[3]{C. D. Han}
\author[2,16]{T. Hashimoto}
\author[4]{M. Iliescu}
\author[12]{K. Inoue}
\author[6]{S. Ishimoto}
\author[1]{K. Itahashi}
\author[2]{M. Iwasaki}
\author[16]{Y. Ma}
\author[2]{R. Murayama}
\author[12,6]{H. Noumi}
\author[7]{H. Ohnishi}
\author[9]{S. Okada}
\author[2]{H. Outa}
\author[10,8]{K. Piscicchia}
\author[12]{A. Sakaguchi}
\author[16,2]{F. Sakuma}
\author[6]{M. Sato}
\author[4]{A. Scordo}
\author[12]{K. Shirotori}
\author[4,8]{D. Sirghi}
\author[4,8]{F. Sirghi}
\author[6]{S. Suzuki}
\author[5]{K. Tanida}
\author[1]{T. Toda}
\author[6]{T. Yamaga}
\author[3]{X. Yuan}
\author[15]{P. Zhang}
\author[3]{Y. Zhang}
\author[11]{H. Zhang}

\affiliation[12]{organization={Research Center for Nuclear Physics (RCNP), The University of Osaka},
  city={Ibaraki},
  postcode={567-0047}, 
  country={Japan}
}
            
\affiliation[2]{organization={RIKEN Pioneering Research Institute, RIKEN},
  city={Wako},
  postcode={351-0198}, 
  country={Japan}
}

\affiliation[3]{organization={Institute of Modern Physics, Chinese Academy of Sciences},
  city={Lanzhou},
  postcode={730000}, 
  country={China}
}

\affiliation[4]{organization={Laboratori Nazionali di Frascati dell’ INFN},
  postcode={I-00044 Frascati},
  country={Italy}
}

\affiliation[14]{organization={Faculty of Nuclear Sciences and Physical Engineering, Czech Technical University in Prague},
  city={Břehová 7},
  postcode={11519}, 
  country={Prague, Czech Republic}
}
  
\affiliation[16]{organization={RIKEN Nishina Center for Accelerator-Based Science, RIKEN},
  city={Wako},
  postcode={351-0198}, 
  country={Japan}
}

\affiliation[6]{organization={High Energy Accelerator Research Organization (KEK)},
  city={Tsukuba},
  postcode={305-0801}, 
  country={Japan}
}

\affiliation[1]{organization={Department of Physics, The University of Osaka},
            city={Toyonaka},
            postcode={560-0043}, 
            country={Japan}
            }

\affiliation[7]{organization={Research Center of Accelerator and Radioisotope Science (RARiS), Tohoku University},
  city={Sendai},
  postcode={982-0826}, 
  country={Japan}
  }
  
\affiliation[9]{organization={Department of Mathematical and Physical Sciences, Chubu University},
  city={Kasugai},
  postcode={487-8501}, 
  country={Japan}
}

\affiliation[10]{organization={Centro Ricerche Enrico Fermi - Museo Storico della Fisica e Centro Studi e Ricerche "Enrico Fermi"},
  postcode={00184 Rome}, 
  country={Italy}
}


\affiliation[8]{organization={Horia Hulubei National Institute of Physics and Nuclear Engineering (IFIN-HH)},
  city={Magurele},
  country={Romania}
}

\affiliation[5]{organization={Advanced Science Research Center, Japan Atomic Energy Agency (JAEA)},
  city={Tokai},
  postcode={319-1195}, 
  country={Japan}
}

\affiliation[15]{organization={School of Physics and Astronomy, Sun Yat-sen University},
  city={Zhuhai},
  postcode={519082}, 
  country={China}
}
  
\affiliation[11]{organization={Lanzhou University},
  city={Lanzhou},
  postcode={730000}, 
  country={China}
}

\begin{abstract}

Light $s$-shell hypernuclei ($^{3,4}_{\Lambda}\text{H}$) and their ground-state properties are crucial benchmarks in hypernuclear physics. 
In particular, comparing the production cross sections of $^{3}_{\Lambda}\text{H}$ and $^{4}_{\Lambda}\text{H}$ provides insights into the $\Lambda N$ interaction in different isospin configurations, which can help address recent discrepancies in the reported $\Lambda$ binding energy of hypertriton.
We present the first measurement of the production cross sections for $^{3}_{\Lambda}\text{H}$ and $^{4}_{\Lambda}\text{H}$ using the in-flight $(K^-, \pi^0)$ reaction at a beam momentum of 1.0 GeV/$c$ with an identical experimental setup.
The production cross sections in the laboratory frame, for the angular range from 0$^{\circ}$ to 20$^{\circ}$, are measured to be $15.0~\pm~2.6~(\text{stat.})~^{+2.4}_{-2.8}~(\text{syst.})~\mu \text{b} $ and $49.9~\pm~2.1~(\text{stat.})~^{+7.8}_{-8.0}~(\text{syst.})~\mu \text{b}$ for the ground-state of $^{3}_{\Lambda}\text{H}$ and $^{4}_{\Lambda}\text{H}$, respectively.
Using the ratio of these cross sections and comparing it with theoretical calculations, we evaluate the $\Lambda$ binding energy of hypertriton, yielding a value consistent with the picture of a loosely bound system.

\end{abstract}




\end{frontmatter}



\section{Introduction}
\label{introduction}

Hypernuclear physics explores the interaction between hyperons and nucleons ($YN$) within nuclei. 
Over the past few decades, both experimental and theoretical studies of hypernuclei have significantly expanded our understanding of $YN$ interaction through systematic investigations of hypernuclear structures \cite{big_paper}.
However, as the cornerstone of hypernuclear physics, the lightest $s$-shell hypernucleus, hypertriton ($^{3}_{\Lambda}\text{H}$), remains not fully understood. 
In particular, the STAR collaboration has reported the $\Lambda$ binding energy of hypertriton ($B_\Lambda = 0.41 \pm 0.12 ~(\text{stat.}) \pm 0.11 ~(\text{syst.}) ~\text{MeV}$) \cite{STAR_binding_energy} that differs from those obtained in an earlier emulsion experiment ($B_\Lambda = 0.15 \pm 0.08	~\text{MeV}$) \cite{emulsion} and the ALICE collaboration ($B_\Lambda = 0.102 \pm 0.063 ~(\text{stat.}) \pm 0.067 ~(\text{syst.})	~\text{MeV}$) \cite{ALICE2023}. 
Recently, new results from emulsion experiments (J-PARC E07) have been published ($B_\Lambda = 0.23 \pm 0.11 ~(\text{stat.}) \pm 0.05 ~(\text{syst.}) ~\text{MeV}$) \cite{E07_binding_energy}.
Such a discrepancy can result in a large lifetime difference because the spreading of the $\Lambda$ wave function in hypertriton sharply depends on degrees of smallness of the $\Lambda$ binding energy and affects the weak decay rate, which may help explain the long-standing hypertriton lifetime puzzle. 
Moreover, as the only known hadronic system that can be classified as an Efimov state, hypertriton presents a unique opportunity to study the $T=0$ three-body force in a $YNN$ system \cite{Efimov}. 
Notably, the ground-state angular momentum of hypertriton has so far only been indirectly inferred as $J=1/2$ based on the branching ratios of the two-body mesonic decay and inclusive mesonic decay channels \cite{Keyes}.

To address these unresolved issues, the J-PARC E73 collaboration has proposed a novel method for hypernuclei production using the in-flight $^A$Z($K^-$, $\pi^0$)$^{A}_\Lambda$(Z-1) reaction at a beam momentum of 1.0 GeV/$c$ \cite{H4L}. 
This method allows us the direct production of $s$-shell hypernuclei in their ground-state due to the small recoil momentum and spin non-flip dynamics \cite{harada}. 
This method is also particularly advantageous for determining the ground-state quantum number and the $\Lambda$ binding energy of hypertriton when combined with measurements of $^{4}_{\Lambda}\text{H}$ as a reference.

In this paper, we present our recent measurements of the production cross sections of $^{3}_{\Lambda}\text{H}$ and $^{4}_{\Lambda}\text{H}$ using the in-flight $(K^-, \pi^0)$ reaction at a beam momentum of 1.0 GeV/$c$ with the same experimental setup. 
By comparing the production cross sections of $^{3}_{\Lambda}\text{H}$ and $^{4}_{\Lambda}\text{H}$, we evaluate the ground-state quantum number and the $\Lambda$ binding energy of hypertriton.

\section{Experimental Setup}

The J-PARC E73 collaboration aims to resolve the hypertriton lifetime puzzle by employing a method distinct from existing heavy-ion-based experiments \cite{e73}. 
In our approach, hypernuclei are produced via the in-flight $(K^-, \pi^0)$ reaction at a forward angle using a 1.0 GeV/$c$ $K^-$ meson beam. 
In this reaction, a $u$-quark in the target proton is replaced by an $s$-quark, thereby converting the participant proton into a $\Lambda$ hyperon. 
This reaction mechanism enables the selective production of hypernuclei with minimal recoil, providing a precise means to measure their lifetime directly.

Unlike the full reconstruction of the $\pi^0$ momentum as performed in Ref. \cite{nms}, the $(K^-, \pi^0)$ reaction in our setup is identified by detecting a high-energy $\gamma$-ray from the $\pi^0$ decay at a forward angle using a compact \v{C}erenkov electromagnetic calorimeter made of PbF$_2$ crystals. 
As demonstrated in our recent measurement of the $^{4}_{\Lambda}\text{H}$ lifetime \cite{H4L}, after combining with the detection of the subsequent mono-energetic pions from $^{3,4}_\Lambda\text{H} \rightarrow  {^{3,4}\text{He}} + \pi^-$ two-body mesonic weak decay (MWD) at rest, we can effectively identify the production of $_\Lambda^{3,4}\text{H}$ hypernuclei. 
For more details, the experimental methods are described in Ref. \cite{H4L}.

\begin{figure}[h]
    \begin{center}
        \includegraphics[width=8.0cm]{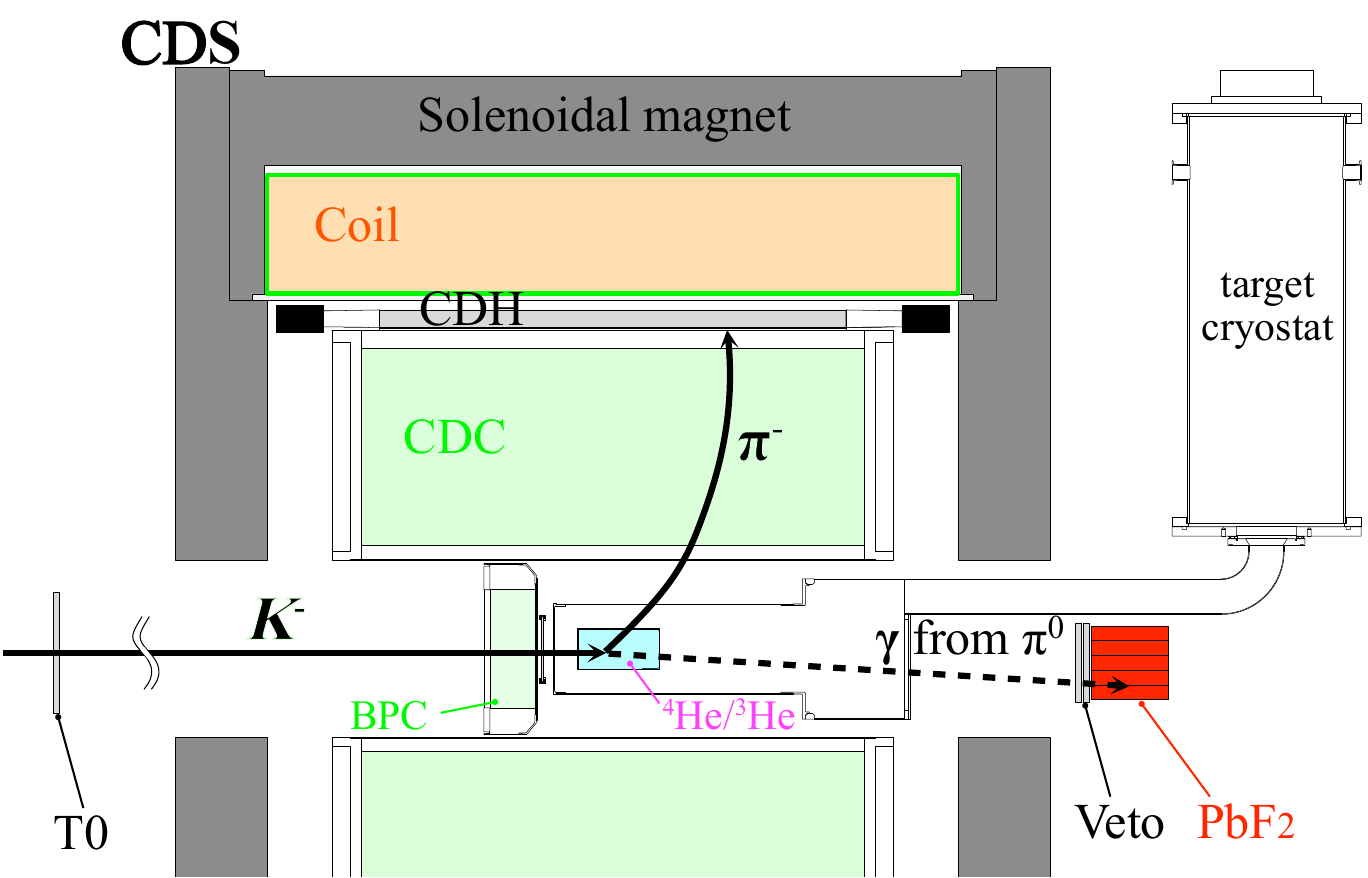}
        \caption{Schematic view of the experimental setup with a timing counter (T0), a Beam Profile Chamber (BPC), a cryogenic system for the liquid $^{3,4}\text{He}$ target;
            high-energy $\gamma$-rays are tagged with a PbF$_2$ calorimeter; 
            Cylindrical Detector System (CDS) is a tracking device to measure $\pi^-$ from the $^{3,4}_\Lambda$H MWD.}
        \label{fig:setup}
    \end{center}
\end{figure}

The schematic view of the experimental setup is shown in Fig. \ref{fig:setup}.
The $K^-$ beam is incident from the left side.
The experimental setup consists of a timing counter (T0), a Beam Profile Chamber (BPC), a cryogenic system for the liquid $^{3,4}\text{He}$ targets, a forward electromagnetic calorimeter, and a Cylindrical Detector System (CDS).
The electromagnetic calorimeter is placed in the forward direction to detect the high energy $\gamma$-rays emitted from the $\pi^0$ decay. 
CDS comprises a solenoidal magnet, a Cylindrical Drift Chamber (CDC), and a Cylindrical Detector Hodoscope (CDH), which is used to detect the $\pi^-$ meson from the two-body MWD $_\Lambda^{3,4}\text{H} \rightarrow {^{3,4}\text{He}} + \pi^-$ to identify the production of the $^{3,4}_\Lambda\text{H}$ hypernucleus. 
Details of CDS can be found in Ref. \cite{e15}.

The results presented in this paper are based on data collected during two pilot runs before the main E73 physics data taking. 
Specifically, the $^{4}_{\Lambda}\text{H}$ production data was acquired in June 2020 at the J-PARC Hadron Experimental Facility's K1.8BR beam line as part of the J-PARC T77 test experiment. 
The $^{3}_{\Lambda}\text{H}$ production data was obtained in May 2021 using the same setup during the first phase of the J-PARC E73 experiment. 
A summary of the data is provided in Table \ref{tab:data_summary}.

\begin{table*}[!h]
    \caption[Data summary]{Data summary for E73 pilot runs.}
    \centering
    \begin{tabular}{cccc}
        \hline\hline
        {Target} & {Hypernucleus} & {Run period}         & $K^-$ beam on target \\
        \hline
        $^4$He   & $^4_\Lambda$H  & 2020/6/20--2020/6/26 & 9.28 $\times 10^9$ \\
        $^3$He   & $^3_\Lambda$H  & 2021/5/11--2021/5/19 & 18.2 $\times 10^9$ \\
        \hline\hline
    \end{tabular}
    \label{tab:data_summary}
\end{table*}

\section{Data analysis}

\subsection{Event selection}

\begin{figure*}[!h]
   \centering
   \includegraphics[width=2.0\columnwidth]{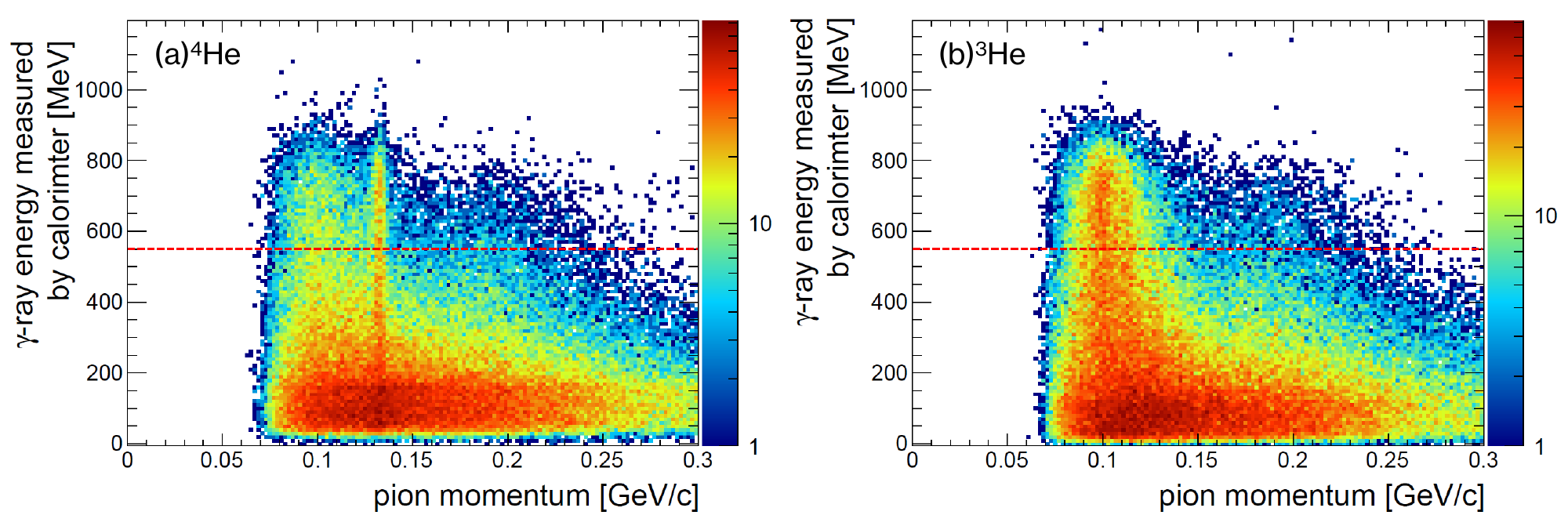}
   \caption[]{
      Contour plots of the $\pi^-$ momentum versus $\gamma$-ray energy measured with the PbF$_2$ calorimeter. 
      (a)~$^4\text{He}$ target. 
      (b)~$^3\text{He}$ target.
      The red horizontal dashed line at 550 MeV indicates the threshold for the event selection.
   }
   \label{fig:pion_mom_ene}
\end{figure*}

Candidates of the in-flight $(K^-, \pi^0)$ reaction events are selected using the $K^-_{beam} \otimes \gamma_{forward}$ trigger condition. 
This trigger condition is defined as a coincidence between kaon beam events, in which beam particles passed through a beam aerogel \v{C}erenkov counter (index = 1.05) without producing a signal, and $\gamma$-ray events, in which the PbF$_2$ calorimeter has a finite energy deposit and a veto counter just in front of it has no signal.
The energy deposit on the PbF$_2$ calorimeter is required to be more than 550 MeV to optimize the signal-to-background ratio of the decay $\pi^-$ momentum spectra in CDS, as demonstrated in Fig. \ref{fig:pion_mom_ene}. 
For a detailed description of the $K^-_{\text{beam}}$ and $\pi^-$ tracking analysis and determination of the PbF$_2$ energy deposit selection condition, please refer to Ref. \cite{H4L}.
Since most $^{3,4}_{\Lambda}\text{H}$ hypernuclei decay at rest with a minimal displacement from the production vertex (typically less than 2 millimeters), we ignore the displacement between the production and decay vertices in the following analyzes.
Thus, a vertex is reconstructed from the $K^-$ beam track and the decayed $\pi^-$ track in CDC using a Distance of Closest Approach (DCA) analysis.
In the analysis, fiducial volume selection for the target region and a DCA cut were applied to ensure reliable vertex reconstruction and to suppress background contributions.

After selecting the $(K^-, \pi^0)$ reaction, the $\pi^-$ momentum spectra can be basically described by two major components: quasi-free production of hyperons ($\Lambda$, $\Sigma$) and the events of MWD of hypernuclei. 
As shown in Fig. \ref{fig:pion_mom_fit_final}, the overall data points can be explained by our decomposition. 
The details of the decomposition will be described in the next section.

\subsection{Decomposition of $\pi^-$ momentum spectra}

\begin{figure*}[!h]
   \centering
   \includegraphics[width=2.0\columnwidth]{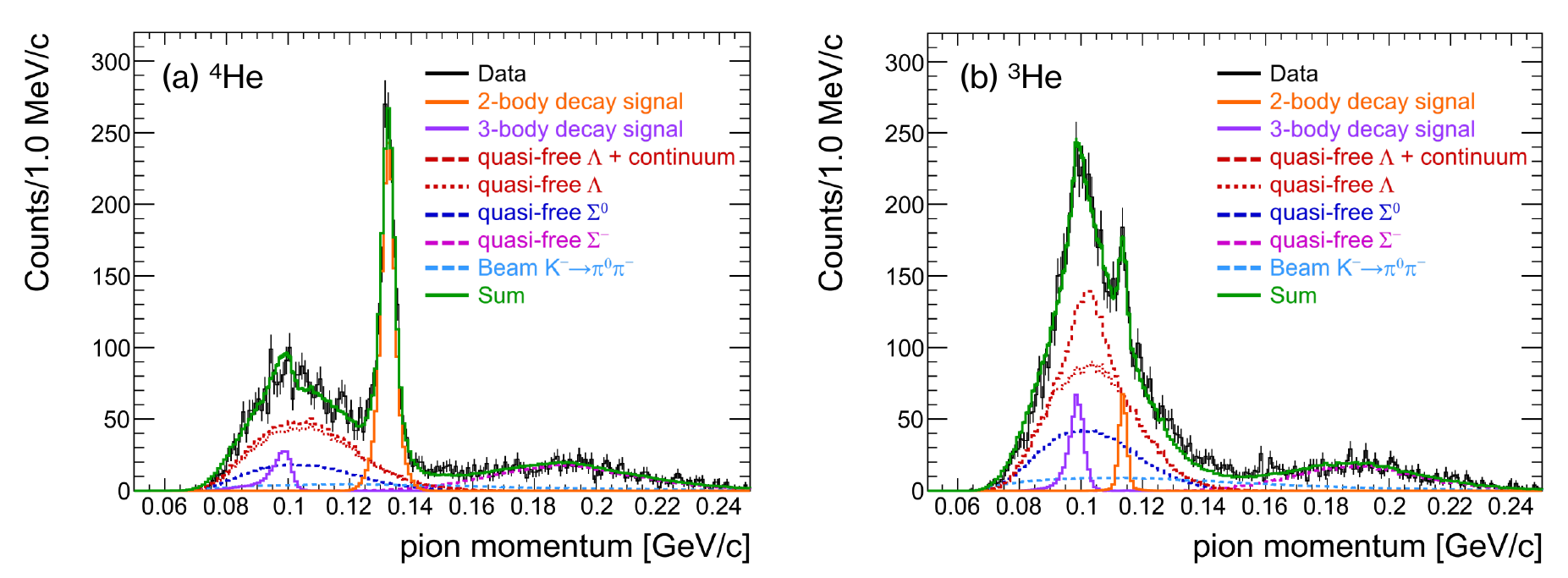}
   \caption[]{
      The $\pi^-$ momentum spectra with fitting components.
      (a)~$^4\text{He}$ dataset.
      (b)~$^3\text{He}$ dataset.
      The black points show the experimental data.
      The orange line denotes the peak from the two-body MWD of hypernucleus while the purple line denotes the peak from the three-body MWD.
      The red dashed line represents the contribution from in-flight $\Lambda$ decay, located in the region 0.07--0.15 GeV/$c$, the blue dashed line from in-flight $\Sigma^0$ decay, located in the same region, the magenta dashed line from in-flight $\Sigma^-$ decay, located in the region 0.15--0.25 GeV/$c$, and the cyan dashed line from beam $K^-$ decay, located in the region 0.07--0.22 GeV/$c$. 
      The green line represents the total spectrum obtained when applying “quasi-free $\Lambda$ + continuum” in the fit.
   }
   \label{fig:pion_mom_fit_final}
\end{figure*}

The number of events used to determine the production cross sections of $_\Lambda^{3,4}\text{H}$ hypernuclei is derived from the $\pi^-$ events produced in the two-body mesonic weak decay (MWD): $_\Lambda^{3,4}\text{H} \rightarrow {^{3,4}\text{He}} + \pi^-$. 
The $\pi^-$ momenta for these decays at rest are approximately 114 MeV/$c$ and 133 MeV/$c$ for $_\Lambda^3\text{H}$ and $_\Lambda^4\text{H}$, respectively. 
To extract the number of events, the peaks are fitted with a Gaussian function, as depicted by the orange lines in Fig.~\ref{fig:pion_mom_fit_final}.
The width ($\sigma$) of the Gaussian peak in the two-body MWD for $_\Lambda^3\text{H}$ around 114 MeV/$c$ is 1.5 MeV/$c$, while the width ($\sigma$) of the Gaussian peak in the two-body MWD for $_\Lambda^4\text{H}$ around 133 MeV/$c$ is 1.7 MeV/$c$.
These values are obtained from the fitting and are consistent with the simulated momentum resolution of the CDC single-track measurement for pions.

The three-body MWD of $_\Lambda^{3,4}\text{H}$ hypernuclei is considered for the decomposition. 
The $\pi^-$ momentum spectra of the three-body MWD of $_\Lambda^{3,4}\text{H}$ are calculated using the theoretical model by Kamada $et~al.$ \cite{kamada1998pi} and Motoba $et~al.$ \cite{motoba1991continuum}, respectively. 
The relative yield between the two-body MWD and the three-body MWD is fixed to the value obtained from previous experimental studies, as 0.357 $^{+0.028}_{-0.027}$ and 0.690 $\pm$ 0.017 for $^3_\Lambda\text{H}$ and $^4_\Lambda\text{H}$, respectively in the Ref. \cite{eckert2021chart}.

For both the $^{3}\text{He}$ and $^{4}\text{He}$ target cases, the $\pi^-$ decayed from quasi-free hyperons constitutes the predominant background in the $\pi^-$ momentum spectra, as illustrated in Fig. \ref{fig:pion_mom_fit_final}. 
This background contribution is estimated using the elementary production cross sections of $\Lambda$ and $\Sigma^{0,-}$ hyperons after convoluted with the Fermi motion of nucleons within the $^{3,4}$He target \cite{jones1975k, conforto1976k}.
Additional background from $K^- \rightarrow \pi^0\pi^-$ decays in-flight is also included in the simulation. 

The final component that needs to be accounted for in the decomposition is the low-momentum $\Lambda$ hyperon originating from the $_\Lambda^{3,4}\text{H}$ continuum states. 
In order to reproduce the observed spectrum, an additional Gaussian component is required. 
This contribution is indispensable in the case of $^3_{\Lambda}$H, while for $^4_{\Lambda}$H it is also included to maintain consistency in the data analysis procedure.
The fit with this continuum component is shown in Fig.~\ref{fig:pion_mom_fit_final}. 
The presence or absence of this continuum component is indicated by the red dashed and dotted lines in the figure, respectively.
It is clear that such a component is enhanced in the $^{3}_\Lambda\text{H}$ spectrum. 
This can be qualitatively understood by the fact that $^4_\Lambda\text{H}$ has several bound states, whereas $^3_\Lambda\text{H}$ has only the ground-state, leading to a relatively larger contribution to the continuum region in the latter case.
It should be noted that the phenomenological decomposition of the $\pi^-$ momentum spectra with the $_\Lambda^{3,4}\text{H}$ continuum states will not degrade the precision of the production cross section measurement, as the signal peaks from the two-body MWD of $_\Lambda^{3,4}\text{H}$ are well separated from the continuum components.

\begin{figure*}[!h]
   \centering
   \includegraphics[width=2.0\columnwidth]{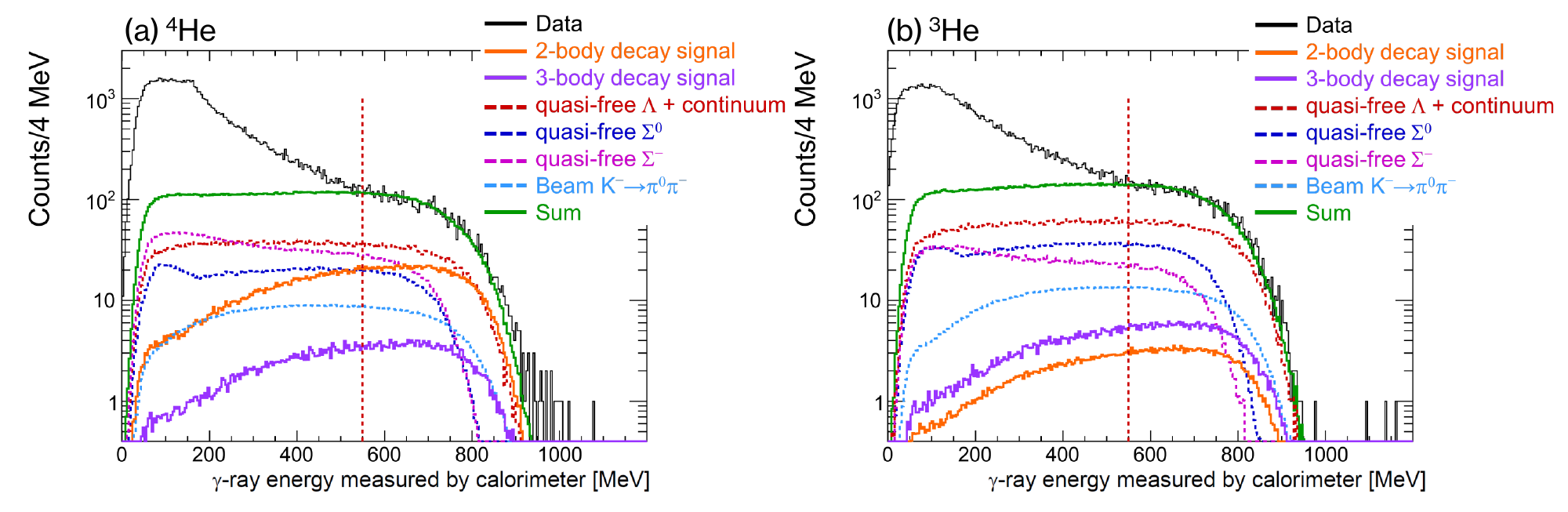}
   \caption[]{
      Distributions of the $\gamma$-ray energy. 
      (a)~$^4\text{He}$ dataset.
      (b)~$^3\text{He}$ dataset.
      Line colors as in Fig.~\ref{fig:pion_mom_fit_final}.
      The vertical red dashed line indicates the selected energy value of 550 MeV.
   }
   \label{fig:calo_energy_fit_final}
\end{figure*}

To further validate the decomposition, we simulate the $\pi^0$ from the $(K^-, \pi^0)$ reaction for each component and compare the PbF$_2$ calorimeter energy distribution with that of the experimental data. 
The results are shown in Fig. \ref{fig:calo_energy_fit_final}. 
The contributions from each component are normalized by the fitting results of the $\pi^-$ momentum spectra. 
The agreement in the spectral shape between the simulation and the experimental data supports the validity of the decomposition in the $E_\gamma \ge 550$ MeV signal region.

After the decomposition, the numbers of $\pi^-$ signals from the two-body MWD of $^{3}_\Lambda\text{H}$ and $^{4}_\Lambda\text{H}$ are estimated to be $225 \pm 39 ~\text{(stat.)} ^{+17}_{-25} ~\text{(syst.)}$ and $1465 \pm 45 ~\text{(stat.)} ^{+5}_{-61} ~\text{(syst.)}$, respectively. 
The statistical uncertainties are derived from the fitting procedure, while the systematic uncertainties are evaluated by varying the binning of the $\pi^-$ momentum spectra, scanning the ratio of $\Lambda$ and $\Sigma^0$ components, and altering the branching ratio of the three-body decay by $1\sigma$ deviation from the adopted values. 
This evaluation is conducted both with and without the inclusion of the continuum component.

\subsection{Acceptance and efficiency correction}

To convert the number of $\pi^-$ events from the two-body MWD to the production cross section of $^{3,4}_{\Lambda}\text{H}$, we consider the detector acceptances, analysis efficiencies, and the two-body MWD branching ratios. 
We evaluate the acceptance and efficiency of CDS and the PbF$_2$ calorimeter using a Monte-Carlo simulation calculation based on the GEANT4 package and apply the same analysis procedures as used for the experimental data.
Table~\ref{tab:cross-section} summarizes the acceptance and efficiencies required to derive the production cross section.

\begin{table*}[!h]
   \centering
   \caption[Summary of the values used to derive the production cross section]{
      Summary of the values used to derive the production cross section.
      Here, $N_\text{2-bodyMWD}$ represents the number of $\pi^-$ events from the two-body MWD detected by CDS, while $L$ denotes the integrated luminosity. 
      The geometrical acceptance of CDS for $\pi^-$ is given by $A_{\text{CDS}}$, and the geometrical acceptance of the PbF$_2$ calorimeter for $\gamma$-rays is indicated by $A_{\text{PbF}_2}$. 
      The $\epsilon$ terms correspond to various analysis efficiencies: $\epsilon_{\text{CDC}}$ is the CDC tracking efficiency, $\epsilon_{\text{PID}}$ is the particle identification efficiency, $\epsilon_{\text{PbF}_2}$ is the energy selection efficiency for $\gamma$-rays, $\epsilon_{\text{DCA}}$ is the selection efficiency for the DCA cut, and $\epsilon_{\text{DAQ}}$ is the DAQ efficiency.
      $Br_{\text{2-body}}$ is the ratio of the two-body mesonic and total weak decay widths of $^{4,3}_{\Lambda}\text{H}$, whose detailed decomposition is listed in Table \ref{tab:Branching_ratio}.
      $R_{g.s./all}$ represents the fraction of ground-state two-body mesonic decays of $^{4,3}_{\Lambda}\text{H}$ among all two-body decay events including those from excited states (see text).
   }
   \begin{tabular}{c | c c c c c c } \hline \hline
      target                     & $^4\text{He}$ &             &                      & $^3\text{He}$ &             &                      \\ \hline
                                 & value    & stat.       & syst.                & value    & stat.       & syst.                \\ \hline
      $N_\text{2-body-MWD}$          & 1465     & $\pm$ 45    & $^{+5}_{-61}$        & 225      & $\pm$ 39    & $^{+17}_{-25}$       \\
      $L$ ($\mu$b$^{-1}$) & 868      & $\pm$ 24    &                      & 984      & $\pm$ 27    &                      \\
      $A_{\text{CDS}}$           & 0.515    &             & $\pm$ 0.003          & 0.498    &             & $\pm$ 0.003          \\
      $A_{\text{PbF}_2}$         & 0.313    &             & $\pm$ 0.002          & 0.319    &             & $\pm$ 0.002          \\ \hline
      $\epsilon_{\text{CDC}}$    & 0.981    & $\pm$ 0.002 &                      & 0.977    & $\pm$ 0.003 &                      \\
      $\epsilon_{\text{PID}}$    & 0.948    &             & $\pm$ 0.003          & 0.965    &             & $\pm$ 0.002          \\
      $\epsilon_{\text{PbF}_2}$  & 0.501    &             & $^{+0.059}_{-0.062}$ & 0.514    &             & $^{+0.060}_{-0.063}$ \\
      $\epsilon_{\text{DCA}}$    & 0.948    &             & $\pm$ 0.005          & 0.919    &             & $\pm$ 0.005          \\
      $\epsilon_{\text{DAQ}}$    & 0.918    & $\pm$ 0.007 &                      & 0.926    & $\pm$ 0.004 &                      \\ \hline 
      $Br_{\text{2-body}}$       & 0.51     &             & $\pm$ 0.05           & 0.23     &             & $\pm$ 0.02           \\ \hline
      $R_{g.s./all}$             & 0.984    &             &                      & 1        &             &                      \\ \hline \hline
   \end{tabular}
   \label{tab:cross-section}
\end{table*}

The CDS acceptance for $\pi^-$ is estimated for the two-body MWD of $^{3,4}_{\Lambda}\text{H}$ hypernuclei, incorporating the theoretical angular distribution from Harada $et~al.$ \cite{harada} for the production process. 
The geometrical acceptance of CDS ($A_{\text{CDS}}$) is derived by dividing the number of $\pi^-$ hits on CDH by the total number of generated hypernuclei, resulting in values of $51.5 \pm 0.3\ (\text{syst.})  ~\%$ for $^4_{\Lambda}$H and $49.8 \pm 0.3\ (\text{syst.}) ~\%$ for $^3_{\Lambda}$H.
Systematic uncertainty due to target position is evaluated as the variation in reconstructed values from a slight shift in nominal position of the target.

The decay pion events are selected with $M^2 \ge 0.015~(\text{GeV/}c^2)^2$ to avoid muon contamination, where the mass squared $M^2$ is determined from tracking and Time-of-Flight analysis. 
The efficiency for $\pi^-$ identification ($\epsilon_{\text{PID}}$) is estimated to be $94.8 \pm 0.3\ (\text{syst.}) ~\%$ for $^4_{\Lambda}$H and $96.5 \pm 0.2\ (\text{syst.}) ~\%$ for $^3_{\Lambda}$H.
Systematic uncertainty from the $M^2$ cutoff is assessed by deviations in computed quantities when altering the cutoff value in the simulation.

The DCA cut of $\le 5$ mm is applied to optimize event selection, resulting in the DCA selection efficiency ($\epsilon_{\text{DCA}}$) of $94.8 \pm 0.5\ (\text{syst.}) ~\%$ for $^4_{\Lambda}$H and $91.9 \pm 0.5\ (\text{syst.}) ~\%$ for $^3_{\Lambda}$H.
Systematic uncertainty from the DCA cutoff is determined by shifts in output parameters due to changes in the cutoff criterion in the simulation.

The geometrical acceptance of the PbF$_{2}$ calorimeter is determined by comparing the total number of simulated $^{3,4}_{\Lambda}\text{H}$ production events with the number of events in which at least a $\gamma$-ray from the $\pi^0$ decay is detected by the PbF$_2$ calorimeter.
The resulting acceptance ratio ($A_{\text{PbF}_2}$) is $31.3 \pm 0.2\ (\text{syst.}) ~\%$ for $^4_{\Lambda}\text{H}$ and $31.9 \pm 0.2\ (\text{syst.}) ~\%$ for $^3_{\Lambda}\text{H}$.
Systematic uncertainty from the z-position of the calorimeter is quantified by fluctuations in measured values from perturbations in its z-direction alignment.

The $\gamma$-ray energy selection efficiencies of the PbF$_2$ calorimeter are determined by comparing the number of simulated $^{3,4}_{\Lambda}\text{H}$ production events with hits in the PbF$_2$ calorimeter with the number of events in which $\gamma$-rays with energies $E_{\gamma} \ge 550$ MeV are detected.
In this experiment, the energy calibration of the calorimeter was performed using 1.0 GeV/$c$ electrons mixed in the meson beam. 
The energy resolution is evaluated to be 5\%, whereas the beam momentum bite is roughly 3\%. 
As there is only one calibration point, ambiguity in determining the calorimeter energy exists. 
This ambiguity is made the uncertainty of the $\gamma$-ray selection efficiencies.
The $\gamma$-ray selection efficiencies of the PbF$_2$ calorimeter ($\epsilon_{\text{PbF}_2}$) are 50.1 $^{+5.9}_{-6.2}$\ (\text{syst.}) ~\% for $^4_\Lambda\text{H}$ and 51.4 $^{+6.0}_{-6.3}$\ (\text{syst.})\% for $^3_\Lambda\text{H}$. 

The total flux of the $K^-$ beam and the integrated luminosity are determined using the scaler count from the kaon beam trigger, the efficiency of the kaon event selection process, and fiducial volume selection in the target. 
The target thickness is 10 cm.
During the period when $^4$He was used, the temperature stayed within a range of 2.83 K to 2.87 K, corresponding to a density of 0.1426 $\pm$ 0.0002 g/cm$^3$.
In the phase of the experiment involving $^3$He, the temperature varied between 2.62 K and 2.71 K, leading to a density of 0.071 $\pm$ 0.001 g/cm$^3$.
The total numbers of $K^-$ particles passing through the fiducial volume are calculated to be (4.03 $\pm$ 0.11\ (\text{stat.})) $\times$ 10$^9$ for the $^4\text{He}$ target and (6.90 $\pm$ 0.17\ (\text{stat.})) $\times$ 10$^9$ for the $^3\text{He}$ target. 
The integrated luminosity is estimated to be 868 $\pm$ 24\ (\text{stat.}) $\mu \rm{b}^{-1}$ and 984 $\pm$ 27\ (\text{stat.}) $\mu \rm{b}^{-1}$ for the $^4\text{He}$ and $^3\text{He}$ targets, respectively.

The tracking efficiency of CDC is estimated using cosmic-ray data collected during off-spill times. 
A special trigger setup records cosmic-ray events when two CDH detectors on opposite sides of CDC are hit, allowing the selection of events passing near the CDC center.  
We determine the CDC tracking efficiency ($\epsilon_{\text{CDC}}$) to be $98.1 \pm 0.2\ (\text{stat.}) ~\%$ for $^4\text{He}$ target runs and $97.7 \pm 0.3\ (\text{stat.}) ~\%$ for $^3\text{He}$ target runs.

The DAQ system efficiency is measured by comparing the total number of recorded events to that of the triggered events. 
The DAQ efficiency ($\epsilon_{\text{DAQ}}$) is $91.8 \pm 0.7\ (\text{stat.}) ~\%$ for $^4_{\Lambda}$H production and $92.6 \pm 0.4\ (\text{stat.}) ~\%$ for $^3_{\Lambda}$H production.

The $_\Lambda^{4}\text{H}$ two-body decay events obtained in this experiment should include contributions from both the $_\Lambda^{4}\text{H}$ ground and excited states.
To extract the ground-state component, we evaluate the contamination from the $_\Lambda^{4}\text{H}$ excited states using the production cross section and angular distribution calculated in Ref. \cite{harada}.
As a result, the contribution from the $_\Lambda^{4}\text{H}$ excited states that decay to the $_\Lambda^{4}\text{H}$ ground-state and subsequently decay through the two-body channel is evaluated to correspond to 1.6~\%.
Therefore, the ratio of directly produced $_\Lambda^{4}\text{H}$ ground-states ($R_{g.s./all}$) is determined to be 98.4~\%.
While the existence of an excited state of $_\Lambda^{3}\text{H}$ remains unconfirmed, all two-body decays of $_\Lambda^{3}\text{H}$ are assumed to originate from the ground-state in this analysis.

After applying the acceptance and efficiency corrections from Table~\ref{tab:cross-section} to the number of $\pi^-$ signal events from the two-body MWD, we can derive the integrated production cross section with branching ratio applied to two-body MWD for $^{3,4}_{\Lambda}\text{H}$ within the angular range from 0$^\circ$ to 20$^\circ$ in the laboratory frame, which is determined by the coverage of the PbF$_2$ calorimeter. 
The cross section is defined as
\begin{gather*}
   \sigma^{\theta_{\text{lab}} = 0^\circ \text{--} 20^\circ} \times Br_{\text{2-body}}~=  \frac{N_{ \text{2-body~MWD}~\cdot~R_{g.s./all} }}{L \cdot A_{\text{CDS}}\cdot A_{\text{PbF}_2} \cdot \epsilon_{\text{total}} }, \\
    \epsilon_{\text{total}} = \epsilon_{\text{CDC}} \cdot \epsilon_{\text{PID}} \cdot \epsilon_{\text{PbF}_2} \cdot \epsilon_{\text{DCA}} \cdot \epsilon_{\text{DAQ}}. \nonumber
\end{gather*}

As a result, the ground-state production cross sections multiplied by the two-body MWD branching ratios are
\begin{gather*}
   \sigma^{\theta_{\text{lab}} = 0^\circ \text{--} 20^\circ}_{^4_{\Lambda}\text{H}} \times Br_{\text{2-body}}(^4_{\Lambda}\text{H})~=~25.3~\pm~1.1~(\text{stat.})~^{+3.0}_{-3.2}~~(\text{syst.})~\mu \rm{b}, \\
   \sigma^{\theta_{\text{lab}} = 0^\circ \text{--} 20^\circ}_{^3_{\Lambda}\text{H}} \times Br_{\text{2-body}}(^3_{\Lambda}\text{H})~=~3.5~\pm~0.6~(\text{stat.})~^{+0.5}_{-0.6}~(\text{syst.})~\mu \rm{b}.
\end{gather*}

Finally, by applying the two-body MWD branching ratios listed in Table~\ref{tab:Branching_ratio}, we obtain the production cross sections:
\begin{eqnarray*}
   \sigma^{\theta_{\text{lab}} = 0^\circ \text{--} 20^\circ}_{^4_{\Lambda}\text{H}}\ &=&\ 49.9\ \pm\ 2.1\ (\text{stat.})\ ^{+7.7}_{-8.0}\ (\text{syst.})\ \mu \rm{b},\\
   \sigma^{\theta_{\text{lab}} = 0^\circ \text{--} 20^\circ}_{^3_{\Lambda}\text{H}}\ &=&\ 15.0\ \pm\ 2.6\ (\text{stat.})\ ^{+2.4}_{-2.8}\ (\text{syst.})\ \mu \rm{b}.
\end{eqnarray*}
The branching ratios in Table~\ref{tab:Branching_ratio} are derived from a combination of available experimental data and theoretical calculations.
The uncertainties quoted for the cross sections include contributions from all sources, including those associated with our analysis and the branching-ratio inputs.

\begin{table*}[!h]
   \centering
   \caption[Summary of the MWD branching ratios]{
      $\Gamma$ denotes the weak decay width.
      $\Gamma_{\pi^{-/0}}$ represent the total mesonic weak decay widths of $^{4,3}_{\Lambda}\text{H} \rightarrow X + \pi^{-/0}$, including both two- and three-body decay modes.
      $\Gamma_{nm}$ denotes the non-mesonic weak decay width of $^{4,3}_{\Lambda}\text{H}$, which includes both $\Lambda n \rightarrow nn$ and $\Lambda p \rightarrow np$ channels.
      $\Gamma(^{4,3}\text{He}+\pi^-)$ and $\Gamma_{all}$ represent the two-body mesonic and total weak decay widths of $^{4,3}_{\Lambda}\text{H}$, respectively.
      The ratio $\Gamma(^{4,3}\rm{He}+\pi^-) / \Gamma_{\rm all}$ is calculated as $ \Gamma(^{4,3}\text{He}+\pi^-)/\Gamma_{\pi^-} \times 1/((\Gamma_{\pi^-}+\Gamma_{\pi^0}+\Gamma_{nm})/\Gamma_{\pi^-})$.
      The values without uncertainties are taken from theoretical calculations, and $\Gamma(^{4,3}\rm{He}+\pi^-) / \Gamma_{\rm all}$ is evaluated without including these theoretical uncertainties.
    } 
   \begin{tabular}{c | c c } \hline \hline
      Fractional branching ratio                       & $^4\text{He}$                                  & $^3\text{He}$                                          \\ \hline
      $\Gamma(^{4,3}\text{He}+\pi^-)/\Gamma_{\pi^-}$   & 0.690 $\pm$ 0.017  \cite{eckert2021chart, BERTRAND197077,Lock:1964bp,Ismail1963,ammar1961} & 0.357 $^{+0.028}_{-0.027}$ \cite{eckert2021chart,PhysRevC.97.054909,KEYES1973,KEYES1970,Keyes1968,Lock:1964bp,ammar1962,Ismail1963,ammar1961} \\
      $\Gamma_{\pi^0}/\Gamma_{\pi^-}$                  & 0.1                \cite{Lock:1964bp}     & 0.5    \cite{kamada1998pi}                        \\
      $\Gamma_{\rm nm}/\Gamma_{\pi^-}$                 & 0.26 $\pm$ 0.13    \cite{Lock:1964bp}     & 0.025  \cite{kamada1998pi}                        \\ \hline
      $\Gamma(^{4,3}\rm{He}+\pi^-) / \Gamma_{\rm all}$ & 0.51 $\pm$ 0.05                           & 0.23 $\pm$ 0.02                                   \\ \hline \hline
   \end{tabular}
   \label{tab:Branching_ratio}
\end{table*}

\section{Discussion}

As a uniquely weakly bound hadronic system ($B_\Lambda: 0.15\sim0.41$ MeV, Ref.~\cite{emulsion, STAR_binding_energy}), the production cross section of $^3_{\Lambda}\text{H}$ is highly sensitive to its $\Lambda$ binding energy. 
Even a small change in the $\Lambda$ binding energy can cause significant variations of the production cross section~\cite{harada}. 
By using the production cross section ratio between $^3_{\Lambda}\text{H}$ and $^4_{\Lambda}\text{H}$, the $\Lambda$ binding energy of $^3_{\Lambda}\text{H}$ can be derived even more reliably both experimentally and theoretically.

From an experimental point of view, the systematic uncertainties of the production cross section ratio are partially canceled out such as the geometrical acceptances of CDS and PbF$_2$ and the efficiencies of detectors. 
In particular, it can significantly cancel the systematic uncertainty caused by energy selection in PbF$_2$ calorimeters.
The theoretical calculation of the production cross section ratio based on the same DWIA framework is also more robust against the ambiguities of the wave functions.

Based on the production cross section obtained in the previous section, the ratio $R_{34}$ = $\sigma_{^3_{\Lambda}\text{H}}/\sigma_{^4_{\Lambda}\text{H}}$ is derived as follows:
\begin{eqnarray*}
    R_{34} = 0.300 \pm 0.054\ (\text{stat.})\ ^{+0.047}_{-0.051}\ (\text{syst.}),
\end{eqnarray*}
where the systematic uncertainty is re-estimated by taking into account the partial cancellation of the systematic uncertainties of the production cross sections.
The systematic uncertainties of the individual production cross section and the $R_{34}$ ratio are summarized in
Table~\ref{tab:Systematic_uncertainties_cross-secttion}.

\begin{table*}[!h]
    \centering
    \caption[Summary of the systematic uncertainties for the production cross sections and $R_{34}$]{
        Summary of the systematic uncertainties for the production cross sections and $R_{34}$.}
    \begin{tabular}{c | c c |c} \hline \hline
        Contribution                                  &   $^4\text{He}$                  &   $^3\text{He}$                      &   $R_{34}$              \\ \hline
        fitting process                               &   $^{+0.2}_{-1.3}$ $\mu$b   &   $^{+1.1}_{-1.7}$ $\mu$b       &   $^{+0.024}_{-0.033}$  \\
        $\gamma-$ray selection in PbF$_2$ calorimeter &   $^{+5.9}_{-6.2}$ $\mu$b   &   $^{+1.7}_{-1.8}$ $\mu$b       &   $^{+0.012}_{-0.005}$  \\
        Acceptance of CDS and PbF$_2$ calorimeter     &   $\pm$ 0.4 $\mu$b          &   $\pm$ 0.2 $\mu$b              &   $\pm$ 0.005           \\
        Analysis efficiency                           &   $\pm$ 0.3 $\mu$b          &   $^{+0.09}_{-0.08}$ $\mu$b     &   $\pm$ 0.002           \\
        Branching ratio                               &   $\pm$ 4.9 $\mu$b          &   $^{+1.2}_{-1.1}$ $\mu$b       &   $^{+0.038}_{-0.037}$  \\ \hline
        Total (quadratic sum)                         &   $^{+7.7}_{-8.0}$ $\mu$b   &   $^{+2.4}_{-2.8}$ $\mu$b       &   $^{+0.047}_{-0.051}$  \\ \hline \hline
    \end{tabular}
    \label{tab:Systematic_uncertainties_cross-secttion}
\end{table*}

\begin{figure}[htbp]
    \centering
    \includegraphics[width=\columnwidth]{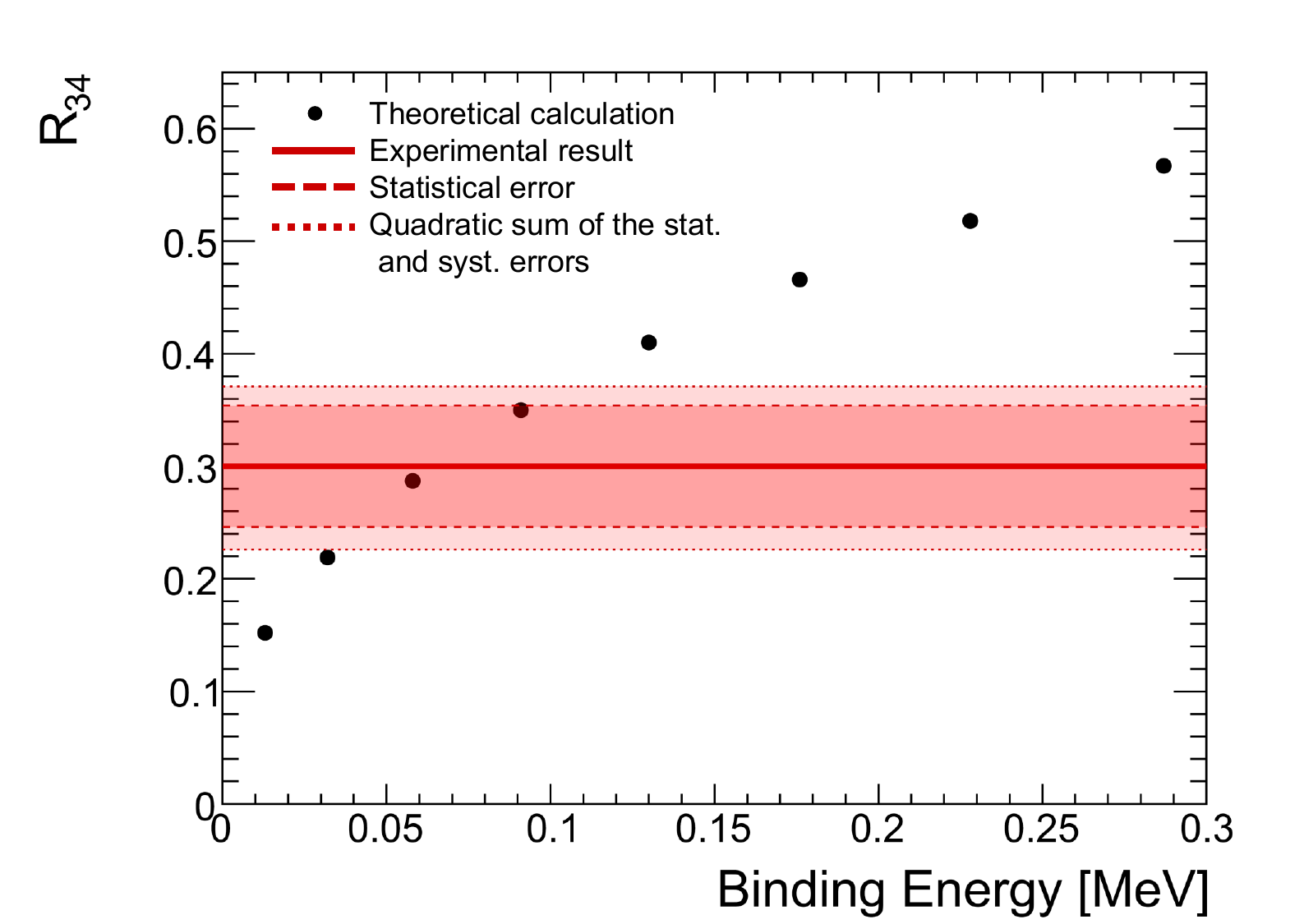}
    \caption{
    The calculated production cross section ratio $R_{34} = \sigma_{^3_{\Lambda}\text{H}}/\sigma_{^4_{\Lambda}\text{H}}$ varying the assumed $\Lambda$ binding energy of hypertriton.
    The horizontal red line indicates the measured ratio $R_{34}$, the long dashed lines show the range of the statistical errors, and the short dashed lines show the total errors obtained as the quadratic sum of the statistical and systematic errors.
    The dots represent the calculated ratios. 
    The theoretical calculation by Harada \cite{harada2023private} is plotted to illustrate the consistency or discrepancy between the experimental result and the theoretical expectation.
    }
    \label{fig:cross_section_ratio}
\end{figure}

Figure~\ref{fig:cross_section_ratio} shows the calculated production cross section ratio varying the assumed $\Lambda$ binding energy of hypertriton.
The horizontal red line in Fig.~\ref{fig:cross_section_ratio} is the value derived from the present work, the long dashed lines show the range of the statistical error, and the short dashed lines show the total errors obtained as the quadratic sum of the statistical and systematic errors.
The theoretical calculation by Harada \cite{harada2023private} is plotted to illustrate the consistency or discrepancy between the experimental result and the theoretical expectation.
The black dots are the theoretical production cross section ratio $R_{34}$. 
By comparing the theoretical calculation with the experimental values, $\Lambda$ binding energy of $^3_{\Lambda}\text{H}$ is estimated as follows:
\begin{eqnarray*}
    B_{\Lambda} = 0.063\ ^{+0.029}_{-0.023}\ (\text{stat.})\ ^{+0.025}_{-0.021}\ (\text{syst.})\ \text{MeV},
\end{eqnarray*}
where the theoretical uncertainty is not included but expected to be most part of the uncertainty would be canceled out when taking the ratio. 
Our result for $B_\Lambda$ is slightly smaller than the emulsion average \cite{emulsion} and the J-PARC E07 result \cite{E07_binding_energy}, but still consistent with them within about $2\sigma$. 
Compared to the STAR measurement \cite{STAR_binding_energy}, our value is significantly smaller, while it is consistent with the ALICE result \cite{ALICE2023} within the quoted uncertainties. 
This overall comparison suggests a tendency toward a longer lifetime for the hypertriton.

Another important property of the hypertriton is its ground-state spin. 
The ratio $R_{34}\sim 1/3$ clearly demonstrates a sizable production cross section of the $^3\text{He}(K^-, \pi^0)^3_{\Lambda}\text{H}$ reaction, which can be understood by the kinematics and the  $\Lambda$ binding energy of $^3_{\Lambda}\text{H}$ ground-state, as discussed in Ref. \cite{harada}. 
Owing to the spin non-flip dynamics of the in-flight $(K^-, \pi^0)$ reaction, we can conclude that the ground-state spin of $^3_{\Lambda}\text{H}$ produced in our reaction is the same as that of the $^3$He target, namely $J=1/2$.

\section{Summary}

This study presents the first measurement of the production cross sections of $^3_\Lambda\text{H}$ and $^4_\Lambda\text{H}$ ground-state hypernuclei using the in-flight $(K^-, \pi^0)$ reaction at a beam momentum of 1.0 GeV/$c$. 
The experiment aims to address discrepancies in the reported binding energies of hypertriton, one of the crucial benchmarks in hypernuclear physics. 
These measurements are conducted using the J-PARC E73 experimental setup, which combines a forward \v{C}erenkov calorimeter to tag the $(K^-,\pi^0)$ reaction and a cylindrical detector system (CDS) to detect decay products.

The production cross sections, for the angular range from 0$^{\circ}$ to 20$^{\circ}$ in the laboratory frame, are measured as $15.0 \pm 2.6\ \text{(stat.)}\ ^{+2.4}_{-2.8}\ \text{(syst.)} \ \mu\text{b}$ for $^3_\Lambda\text{H}$ and $50.7 \pm 2.1\ \text{(stat.)}\ ^{+7.8}_{-8.3}\ \text{(syst.)}\ \mu\text{b}$ for $^4_\Lambda\text{H}$. 
By comparing the cross section ratio $\sigma(^3_\Lambda\text{H})/\sigma(^4_\Lambda\text{H})$ with theoretical calculations, we evaluated the $\Lambda$ binding energy of the hypertriton, which yielded a value consistent with the picture of a loosely bound system.

This result provides a determination of the $\Lambda$ binding energy of hypertriton using an alternative, theoretically grounded approach, which is essential for resolving the hypertriton lifetime puzzle and for understanding the $\Lambda N$ interaction in light hypernuclei.
Furthermore, the observed production cross section ratio $R_{34} \sim 1/3$, combined with the spin non-flip nature of the $(K^-,\pi^0)$ reaction, supports that the ground-state spin of $^3_\Lambda\text{H}$ is $J = 1/2$. 
Ongoing data analysis with the newly obtained data from February 2025 will focus on further refining the hypertriton lifetime measurement, which is expected to shed light on the extensively debated puzzle in the hypernuclear physics.

\section*{Acknowledgements}
We would like to express our appreciation for the support from the J-PARC accelerator and hadron facility staff. 
We are also grateful for the inspiring discussion and kind help from Prof. T. Harada. 
This project is partially supported by the MEXT Grants-in-Aid 17H04842, 19J20135, 21H00129, 18H05402, 22H04917, and 26287057. 
We also acknowledge the support by International Partnership Program of the Chinese Academy of Sciences. Grant No. 016GJHZ2022054FN and The National Natural Science Foundation of China, Grant No. 12305127.
TH was supported by a MEXT Leading Initiative for Excellent Young Researchers Grant. 


\bibliographystyle{elsarticle-num}
\bibliography{citelist.bib}

\end{document}